\documentclass[11pt, a4paper]{article}

\usepackage[utf8]{inputenc} 
\usepackage[T1]{fontenc}    
\usepackage{hyperref}       
\usepackage{url}            
\usepackage{booktabs}       
\usepackage{amsfonts}       
\usepackage{nicefrac}       
\usepackage{microtype}      
\usepackage{lipsum}
\usepackage{graphicx}
\usepackage{amsmath}
\usepackage{authblk}
\usepackage{titlesec}
\titleformat*{\section}{\normalsize\bf}

\title{Spontaneous emission noise in mode-locked lasers and frequency combs}

\author[1,2]{Ruoyu Liao} 
\author[1]{Chao Mei}
\author[2]{Youjian Song}
\author[3,4]{Ayhan Demircan}
\author[,1,5]{G\"unter Steinmeye \thanks{Corresponding author: steinmey@mbi-berlin.de}}

\affil[1]{Max Born Institute for Nonlinear Optics and Short Pulse Spectroscopy
  Max-Born-Stra\ss e 2a, 12489 Berlin, Germany}
  
 \affil[2]{Ultrafast Laser Laboratory, Key Laboratory of Opto-electronic Information Technology, Ministry of Education, School of Precision Instrument and Opto-electronics Engineering, Tianjin University, Tianjin 300072, China}
 
 \affil[3]{Cluster of Excellence PhoenixD,
   Welfengarten 1, 30167, Hannover, Germany}
  \affil[4]{ Institute of  Quantum Optics,
    Leibniz University Hannover,
    Welfengarten 1 30167, Hannover, Germany}
  
  \affil[5]{Institut f\"ur Physik, Humboldt Universit\"at zu Berlin,
  Newtonstra\ss e 15, 12489 Berlin, Germany}
\begin{document}
\maketitle

\begin{abstract}
Amplified spontaneous emission (ASE) causes fluctuations of pulse energy, of the optical phase
and of the timing of the pulse intensity envelope in a mode-locked laser or frequency comb. Starting from the assumption of one ASE photon per longitudinal laser mode and roundtrip, we rederive analytic equations for the three fundamental types of quantum noise in a laser. To this end, we analyze the interference of the coherent intracavity field and a spectrally localized ASE photon as a function of wavelength and phase of the latter. Performing an integration over all wavelengths and phases and taking stochastic noise into account, we compute ASE-induced jitters for all quantities considered. Continuing this approach, we then derive an expression for the resulting carrier-envelope phase noise of the comb, for which so far only numerical estimates exist. We further compute analytical estimates for ASE induced pulse chirp and duration variations and address the issue of resulting pulse contrast in a mode-locked laser and the resulting coherence properties. Considering three example cases, we finally compute estimates for all quantities analyzed. Taken together, our analysis provides a comprehensive view of ASE effects in a mode-locked laser, which unites numerous scattered reports across the literature.
\end{abstract}

\section{Introduction}
Amplified spontaneous emission (ASE) gives rise to a number of noise mechanisms and resulting limitations of laser technology. As originally suggested by Schawlow and Townes, this phenomenon can be understood as random interference of spontaneously emitted photons with the coherent light field propagating inside the laser cavity \cite{Schawlow}. From Einstein's quantum theory of radiation \cite{Einstein,Demtroeder}, one expects that one photon is spontaneously emitted into every mode per cavity roundtrip of a laser, and the spontaneously emitted number of photons is typically large in case of a mode-locked laser as it operates on many modes. As these photons may both increase or decrease the instantaneous laser frequency, phase or frequency fluctuations emerge as a consequence of ASE. The latter effect is known as Schawlow-Townes noise \cite{Schawlow, Paschotta2} and gives rise to the fundamental quantum-limited linewidth of a laser, which scales inversely proportional to the number of intracavity photons. As these fluctuations are often in the sub-hertz range, they are typically not immediately accessible to measurements. Schawlow-Townes noise also manifests itself in a limited coherence length of a continuously operated laser. For sub-hertz linewidths, resulting coherence lengths are in the range of millions of kilometers. Therefore, for a rather large class of lasers, technical noise sources dominate over quantum noise. However, direct observation of the Schawlow-Townes linewidth and coherence length is possible near the laser threshold \cite{Wellegehausen} and for diode lasers with their rather short cavities and modest intracavity powers \cite{Ho,Henry,Rush}. Another obvious consequence of these interference effects are power fluctuations of the laser, that is, an effect which is widely known as shot noise \cite{Rice,Paschotta4,Ho}.  In mode-locked lasers, an additional quantum noise effect appears, giving rise to an ASE induced timing jitter of the pulses \cite{HausMecozzi,Paschotta,Kim}. In the absence of intracavity coupling mechanisms, Schawlow-Townes noise and the timing jitter are uncorrelated and result in fluctuations of the carrier-envelope phase, i.e., the relative phase between the peak of the intensity envelope of a pulse and the underlying electric-field carrier. While analytic expressions for the Schawlow-Townes noise and the timing jitter of a pulse are known, carrier-envelope phase noise has so far only been numerically simulated \cite{Paschotta,Fuji,Fortier}. Here we derive a completely analytic expression for the carrier-envelope phase noise under assumption of Gaussian pulse shape and spectrum. Moreover, we present analytic computations for ASE induced variations of pulse duration and chirp. In a mode-locked laser, ASE also gives rise to the formation of a pedestal that limits pulse contrast. The appearance of this pedestal is accompanied by a degradation of the coherence of the laser. Among all discussed implications of ASE noise, pulse contrast is probably the most directly measurable effect. All other ASE-induced effects are weak for solid-state lasers with large intracavity powers and cavity length. However, quantum noise may cause sizeable fluctuations, for example, of pulse duration in semiconductor lasers, which may also explain some of the experimentally observed difficulties in obtaining stable mode-locking of these promising laser sources.

\section{SEMI-CLASSICAL GAUSSIAN MODEL}
\begin{figure}
    \centering
    \includegraphics[width=0.6 \linewidth]{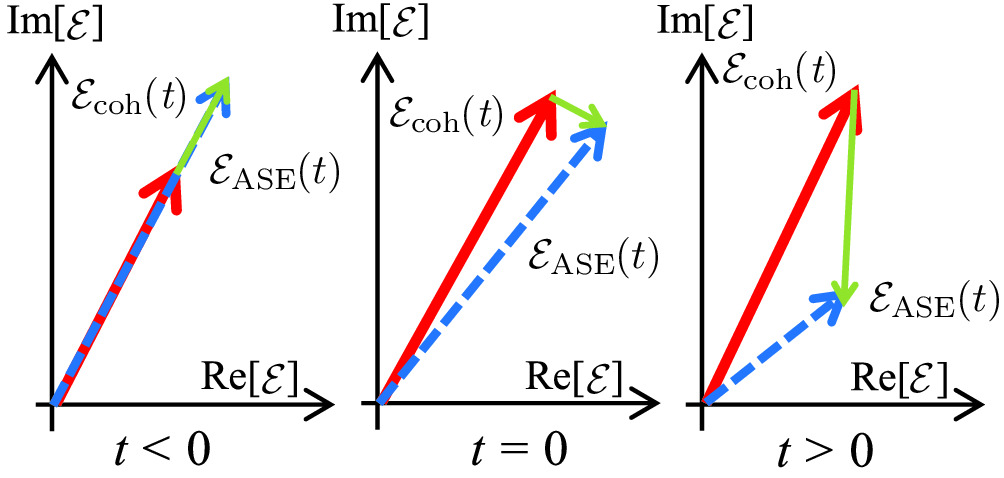}
    \caption{Visualization of the semi-classical model \cite{Henry} applied to the case of a short-pulse laser. The coherent field $\mathcal{E}_{\rm coh}(t)$ interferes with the continuous incoherent field $\mathcal{E}_{\rm ASE}(t)$ representing ASE. For the scenario shown, interference is constructive in the leading edge of the pulse ($t < 0 $) and destructive in the trailing edge  ($t > 0 $). This effectively causes an advance of the pulse waveform relative to undisturbed case.}
    \label{figure1}
\end{figure}

Our semi-classical approach follows the original idea of Charles
Henry \cite{Henry}, representing the elementary ASE effect by vector additions of a coherent complex electric-field waveform $\mathcal{E}_{\rm coh}$(\textit{t}) and an incoherent field $\mathcal{E}_{\rm ASE}$(\textit{t}) of individual ASE photons, see Fig.~$\ref{figure1}$. Using this approach, we first explore the effect of an ASE field with random phase and frequency on the major observables, i.e., energy, phase, and timing. Integrating over all possible phases and frequencies, we then derive expectation values for the pulse-to-pulse jitters of these observables. Here spontaneous emission adds $M$ individual random photons per roundtrip to the circulating coherent field, where $M$ is the effective number of longitudinal modes within the gain bandwidth of the laser. In the semi-classical picture of Henry, the photon concept is of purely statistical nature, i.e., one cannot ascribe a pulse duration to an individual photon. Neither does energy conservation hold on the single-photon level \cite{Pollnau}. Using Gaussian statistics, fluctuations of laser parameters follow a $\sqrt{M}$ relationship, i.e., we can statistically describe the net effect of the $M$ random photons per roundtrip by a properly scaled field amplitude. In the following, we demonstrate that this approach leads to the standard shot-noise limit as well as to the findings of Schawlow and Townes. In particular, the fields associated with the individual photons are spectrally localized, yet have no temporal localization. In this picture, the effect of shot noise is rather intuitive, that is, an added photon may either constructively or destructively interfere with $\mathcal{E}_{\rm coh}(t)$. Similarly,
frequency noise also appears easy to understand as ASE photons may be emitted either above or below the laser center frequency $\omega_0$, causing a pull of the latter to the blue or red, respectively. However, this picture originates from the continuous-wave laser and is probably too simplistic to describe a mode-locked laser. For example, an ASE photon may well be emitted on the red side of the spectrum, yet can interfere constructively
on the blue side of the spectrum and therefore cause an
effective blue shift of the spectrum. Similar considerations can be made in the time domain, i.e., depending on whether a photon predominantly interferes constructively or destructively in the leading edge of the repetitive waveform, giving rise to an advance or delay of the waveform, respectively. In the case shown in Fig.~$\ref{figure1}$, interference is constructive in the leading edge, effectively advancing the propagation of the waveform relative to the undisturbed case. At the same time, the effect on the center frequency is nearly negligible, causing only a minor spectral blue shift. As we will show in the following, ASE-induced spectral and temporal shifts as well as pulse energy  fluctuations are completely uncorrelated in our semi-classical picture.

Let us assume a mode-locked laser with repetition rate $f_{\rm rep}$, intracavity pulse energy $E_{\rm p}$, a Gaussian envelope with duration $\tau$, giving rise to a field envelope
\begin{equation} \label{5}
    \mathcal{E}_{\rm coh}(\textit{t})=\hat{\mathcal {E}} \exp \left(-\frac{\textit{t}^2}{2\tau^2}\right),
\end{equation}
where $\hat{\mathcal{E}}$ is the peak electric field amplitude. We further assume for simplicity that the absolute phase of the pulse is initially zero. Let us define pulse fluence $F_{\rm p}$ and energy $E_{\rm p}$ according to
\begin{equation}\label{6}
    F_{\rm p}=\frac{E_{\rm p}}{A_{\rm eff}}=\frac{\sqrt{\pi}}{Z_0}\hat{\mathcal{E}}^2\tau,
\end{equation}
where $A_{\rm eff}$ is the effective mode area, $\textit{Z}_0 = 1/\textit{c}\epsilon_0 \approx 377\ \Omega$ is the vacuum impedance, and $\textit{c}$ and $\epsilon_0$ are the speed of light and permittivity in vacuum, respectively.

Given the photon energy $E_{\phi} = \hbar\omega_{0}$, we can relate to the average number of photons per pulse $N = E_{\rm p}E_{\phi}$. Let us treat an individual ASE photon as temporally delocalized and define resulting fluence $F_{\phi}=E_\phi/A_{\rm eff}$ and field strength
\begin{equation} \label{6a}
\mathcal{E}_\phi = \sqrt{Z_0 F_\phi f_{\rm rep}}.
\end{equation}
Rewriting Eq.~(\ref{6}) then yields an expression for the peak field strength
\begin{equation}\label{7}
    \hat{\mathcal{E}}=\sqrt{\textit{Z}_0\frac{N F_{\phi}}{\sqrt{\pi}\tau}}.
\end{equation}

\section{INTERFERENCE EFFECT ON PULSE OBSERVABLES}
Let us further assume in our semi-classical approach that the circulating coherent field $\mathcal{E}_{\rm coh}(\textit{t})$ interferes with a
random continuous-wave field
\begin{equation} \label{8}
    \mathcal{E}_{\rm ASE}(t)=\sqrt{M} \mathcal{E}_\phi  \exp (i\omega_\phi\textit{t}+i\varphi),
\end{equation}
where $\mathcal{E}_\phi$ is the equivalent field amplitude of an individual ASE photon. Following random statistics, the incoherent field amplitude grows with $\sqrt{M}$ of the spontaneously emitted photons, resulting in a net amplitude $\sqrt{M} \mathcal{E}_\phi$ and phase $\varphi$ of the continuous wave.
Interference with $\mathcal{E}_{\rm ASE}(t)$ from Eq.\ref{5} then results in the total field
 $   \mathcal{E}_{\rm total}(t)=\mathcal{E}_{\rm coh}(t)+\mathcal{E}_{\rm ASE}(t).$
The energy change after interference is
\begin{equation}\label{11}
\begin{split}
     \delta E_{\rm p} &= \frac{1}{Z_0}\int_{-\infty}^{\infty}\left|\mathcal{E}_{\rm total}(t)\right|^2 - \left|\mathcal{E}_{\rm coh}(t)\right|^2      \text{d}\textit{t} \\&
     \approx\frac{1}{Z_0}\int_{-\infty}^{\infty}2\hat{\mathcal{E}}\mathcal{E}_{\phi}\exp\left(-\frac{t^2}{2\tau^2}\right)
     \cos(\omega\text{t}+\varphi)\text{d}\textit{t}\\&
     =2\sqrt{2}E_{\text{p}}\gamma\cos(\varphi)\exp\left(-\frac{\tau^2\omega^2}{2}\right).
\end{split}
\end{equation}
Here we defined the field ratio
\begin{equation}\label{11a}
\gamma = \sqrt{M} \mathcal{E}_{\phi}/\hat{\mathcal{E}}
\end{equation}
 and exploited that $\gamma \ll 1$. Therefore, we can neglect a temporally unlocalized term of the order $M \mathcal{E}_{\phi}^2$. While this apparent violation of energy conservation may be debatable in a continuous-wave laser \cite{Pollnau}, most of the energy of the incoherent field is localized well outside the width of the coherent pulse. Proper mode-locking provided, the ratio of roundtrip time and pulse duration is on the order of $M$ itself. Therefore, neglection of the $\mathcal{E}_{\phi}^2$ term leads to discrepancies that amount to a few individual photon energies. A much larger effect is the formation of a pedestal, which limits pulse contrast and is treated in Section \ref{sec:pedestal}.

\begin{figure}[!hbt]
    \centering
    \includegraphics[width=0.6 \linewidth]{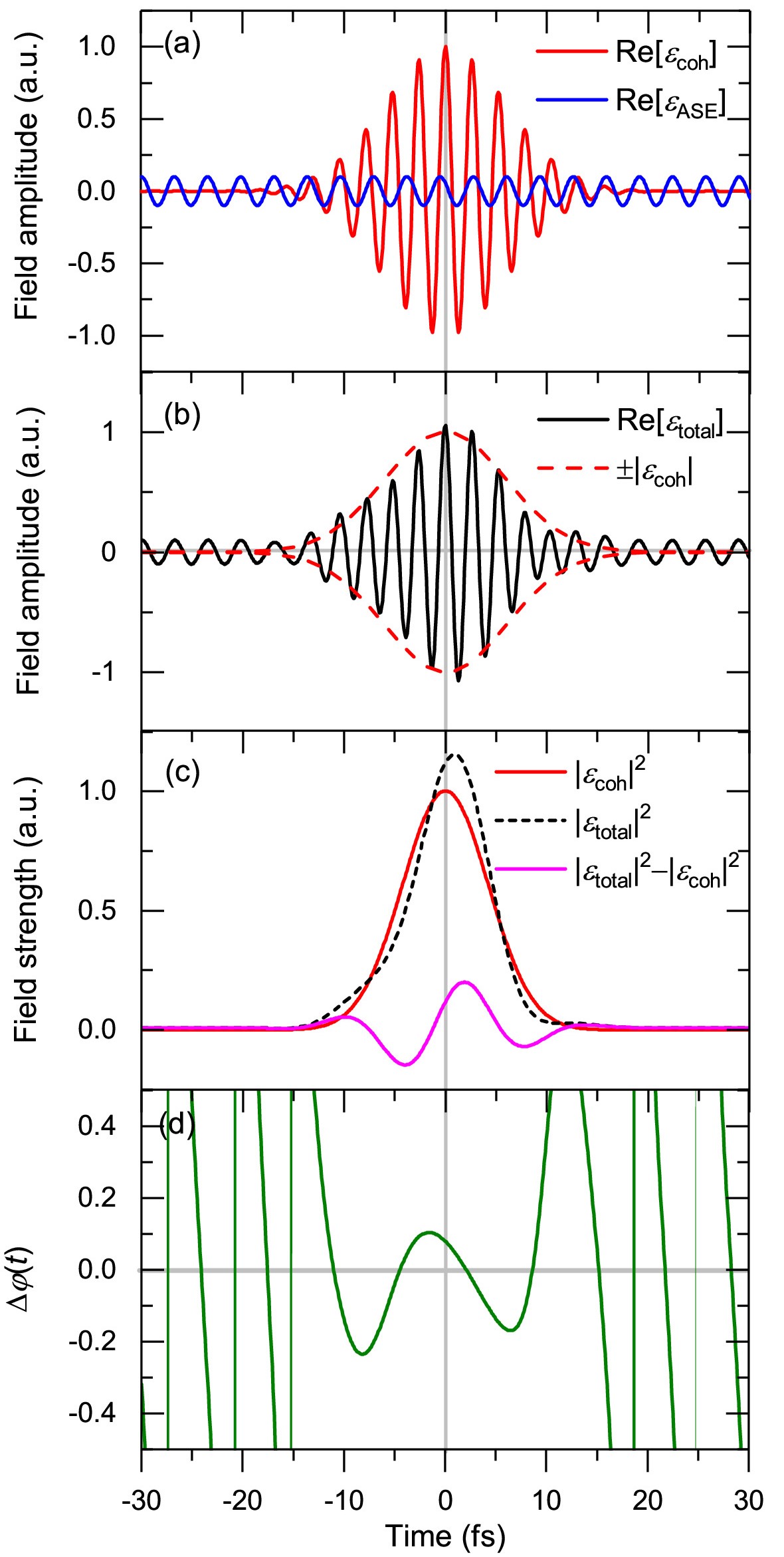}
    \caption{(a) Individual amplitude of Gaussian (red) and continuous (blue) fields. (b) Field amplitude after interference (black solid curve) and envelope of Gaussian field (red dashed curve). (c) Resulting intensity envelope (black short dashed curve) in comparison to original Gaussian envelope (red solid curve) and their intensity difference (pink solid curve ).(d) Phase difference; Schawlow-Townes noise manifests itself as the phase difference at zero delay.}
    \label{figure2}
\end{figure}

In the same fashion as pulse energy, we can evaluate timing jitter
\begin{equation}\label{12}
\begin{split}
    \delta t&=\frac{\int_{-\infty}^{\infty}t\left|\mathcal{E}_{\rm total}(t)\right|^2\text{d}t}{\int_{-\infty}^{\infty}\left|\mathcal{E}_{\rm total}(t)\right|^2\text{d}t}\\&
    \approx\frac{\int_{-\infty}^{\infty}t\left[\exp\left(-\frac{t^2}{\tau^2}\right)+2\gamma\exp\left(-\frac{t^2}{2\tau^2}\right)\cos(\omega\text{t}+\varphi)\right]\text{d}\textit{t}}{\int_{-\infty}^{\infty}\left[\exp\left(-\frac{t^2}{\tau^2}\right)+2\gamma\exp\left(-\frac{t^2}{2\tau^2}\right)\cos(\omega\text{t}+\varphi)\right]\text{d}\textit{t}}\\&
    =\frac{-2\sqrt{2}\gamma\tau^2\omega\sin\varphi\exp\left(-\frac{\tau^2\omega^2}{2}\right)}{1+2\sqrt{2}\gamma\cos\varphi\exp\left(-\frac{\tau^2\omega^2}{2}\right)}.
\end{split}
\end{equation}
Note that $\gamma$ is typically much smaller than unity, so Eq.~$\eqref{12}$ can be simplified as
\begin{equation}\label{13}
    \delta t\approx -2\sqrt{2}\gamma\tau^2\omega\sin\varphi\exp\left(-\frac{\tau^2\omega^2}{2}\right).
\end{equation}
Finally, it is straightforward to compute the resulting phase changes within the pulse profile, which amount to
\begin{equation}\label{14}
    \Delta\varphi(t)=\varphi_{\rm total}-\varphi_{\rm coh}=\arctan \frac{\text{Im}\left[\mathcal{E}_{\rm total}(t)\right]}{\text{Re}\left[\mathcal{E}_{\rm total}(t)\right]}-\omega_0t,
\end{equation}
where $\varphi_{\rm total}$ and $\varphi_{\rm coh}$ are the absolute phase of total field and Gaussian input field, respectively. Im$[\,]$ and Re$[\,]$ represents the imaginary and real parts, respectively. Substituting Eqs.~$\eqref{5}$ and $\eqref{8}$ into Eq.~$\eqref{14}$, we yield
\begin{equation}\label{15}
    \Delta\varphi(t)=\text{arctan}\frac{\exp\left(-\frac{t^2}{2\tau^2}\right)\sin\omega_0t+\gamma\sin(\omega\textit{t}+\varphi)}
    {\exp\left(-\frac{t^2}{2\tau^2}\right)\cos\omega_0t+\gamma\cos(\omega\textit{t}+\varphi)}-\omega_0t.
\end{equation}
For the field components of ultrashort pulse near the center angular frequency, $\omega_0\textit{t} \approx 0$. In this scenario, Eq.~$\eqref{15}$ can be reduced to
\begin{equation}\label{16}
    \Delta\varphi(t)\approx\frac{\gamma\sin(\omega\textit{t}+\varphi)\exp\left(\frac{t^2}{2\tau^2}\right)}{1+\gamma\cos(\omega\textit{t}+\varphi)\exp\left(\frac{t^2}{2\tau^2}\right)}
\end{equation}
when the ratio of the interfering fields is small. This expression can now be used for computing the effective phase change by integrating over the pulse profile
\begin{equation}\label{17}
    \delta\varphi=\frac{\int_{-\infty}^{\infty}\Delta\varphi(t)|\mathcal{E}_{\rm total}(t)|^2\text{d}t}{\int_{-\infty}^{\infty}|\mathcal{E}_{\rm total}(t)|^2\text{d}t}.
\end{equation}
Substituting Eqs.~$\eqref{5}$, $\eqref{8}$ and $\eqref{16}$ into Eq.~\eqref{17} and neglecting terms of order $\mathcal{E}_{\phi}^{2}$ and $\gamma^2$,  we reach
\begin{equation}\label{18}
\begin{split}
    \delta\varphi&\approx\frac{\int_{-\infty}^{\infty}\gamma\sin(\omega\textit{t}+\varphi)\exp\left(-\frac{t^2}{2\tau^2}\right)\text{d}t}{\int_{-\infty}^{\infty}\exp\left(-\frac{t^2}{\tau^2}\right)+2\gamma\exp\left(-\frac{t^2}{2\tau^2}\right)\cos\left(\omega\textit{t}+\varphi\right)\text{d}t}\\&
    =\frac{\sqrt{2}\gamma\sin\varphi\exp\left(-\frac{\tau^2\omega^2}{2}\right)}{1+2\sqrt{2}\gamma\cos\varphi\exp\left(-\frac{\tau^2\omega^2}{2}\right)}\\&
    \approx\sqrt{2}\gamma\sin\varphi\exp\left(-\frac{\tau^2\omega^2}{2}\right).
\end{split}
\end{equation}
The computations in this section are visualized in Fig.~$\ref{figure2}$ where the interference of one ASE photon with a temporally localized wavepacket is demonstrated. Rather large values, i.e., a field amplitude ratio of 0.1, a phase difference of 1\,rad and a frequency difference of 20$\%$ have been chosen for clarity. In particular, as shown in Figs. $\ref{figure2}$(c) and  $\ref{figure2}$(d), the effect of the perturbation field on the intensity scales proportional to pulse envelope whereas the phase effect scales inversely proportional to the envelope. The latter behavior is counteracted by the intensity proportional Gaussian factor, i.e., the pulse phase change is effectively dominated by the behavior near zero delay. Figure  $\ref{figure2}$(c) then also illustrates the effect on the center of gravity of the wavepacket. It is plain to see that the center of gravity experienced a shift towards positive delays, giving rise to the timing jitter of the pulses.
\indent Using the same formalism, we can also evaluate the influence of quantum noise on other pulse observables, e.g., the duration of the intensity envelope
\begin{equation}\label{19}
    \tau=\sqrt{\frac{2 \int_{-\infty}^{\infty}t^2 |\mathcal{E}^2(t)|\,\text{d}t}{\int_{-\infty}^{\infty}|\mathcal{E}^2(t)|\,\text{d}t}}.
\end{equation}
Relating the pulse duration before and after interference with ASE, we compute the change in pulse duration
\begin{equation}\label{20}
\begin{split}
    \delta\tau&=\sqrt{\frac{2 \int_{-\infty}^{\infty}(t-\delta\textit{t})^2|\mathcal{E}_{\rm total}^2(t)|\,\text{d}t}{\int_{-\infty}^{\infty}|\mathcal{E}_{\rm total}^2(t)|\,\text{d}t}}-\sqrt{\frac{2 \int_{-\infty}^{\infty}t^2|\mathcal{E}_{\rm coh}^2(t)|\,\text{d}t}{\int_{-\infty}^{\infty}|\mathcal{E}_{\rm coh}^2(t)|\,\text{d}t}}\\&
    \approx\sqrt{\frac{2 \int_{-\infty}^{\infty}(t^2-2t\delta\textit{t})|\mathcal{E}_{\rm total}^2(t)|\,\text{d}t}{\int_{-\infty}^{\infty}|\mathcal{E}_{\rm total}^2(t)|\,\text{d}t}}-\tau.
\end{split}
\end{equation}
Note that $|\mathcal{E}_{\rm total}(\textit{t})|$ is an even function.
As a result,
\begin{equation}\label{21}
\begin{split}
    \delta\tau &\approx\sqrt{\frac{2 \int_{-\infty}^{\infty}t^2|\mathcal{E}^2_{\rm total}(t)|\,\text{d}t}{\int_{-\infty}^{\infty}|\mathcal{E}_{\rm total}^2(t)| \, \text{d}t}}- \tau \\&
    \approx\tau\bigg[4 \gamma\cos\varphi(1-\tau^2\omega^2)\exp\left(-\frac{\tau^2\omega^2}{2}\right)\\&
    \quad -8 \sqrt{2} \gamma^2\tau^2\omega^2\sin^2\varphi\exp(-\tau^2\omega^2)\bigg]^{\frac{1}{2}}.
\end{split}
\end{equation}
Finally, we can also address quantum-noise induced changes of the chirp
\begin{equation} \label{22}
    \begin{split}
        \beta(t)&=\frac{\partial^2\Delta\varphi(t)}{\partial\textit{t}^2}\\&
        \approx\frac{\gamma\sin(\omega\textit{t}+\varphi)\left(\frac{t^2}{\tau^2}-\tau^2\omega^2+1\right)+2\gamma\omega\textit{t}\cos(\omega\textit{t}+\varphi)}{\tau^2\left[\exp\left(-\frac{t^2}{2\tau^2}\right)+2\gamma\cos(\omega\text{t}+\varphi)\right]}.
    \end{split}
\end{equation}
From this expression, we compute
\begin{equation}\label{23}
    \begin{split}
        \delta\beta&=\sqrt{\frac{\int_{-\infty}^{\infty}\beta(t)|\mathcal{E}_{\rm total}(t)|^2\,\text{d}t}{\int_{-\infty}^{\infty}|\mathcal{E}_{\rm total}(t)|^2\,\text{d}t}}\\&
        \approx\frac{2\sqrt{2}\gamma(\frac{1}{\tau^2}-2\omega^2)\exp\left(-\frac{\tau^2\omega^2}{2}\right)\sin\varphi}{1+2\sqrt{2}\gamma\exp\left(-\frac{\tau^2\omega^2}{2}\right)\cos\varphi}\\&
        \approx2\sqrt{2}\gamma\left(\frac{1}{\tau^2}-2\omega^2\right)\exp\left(-\frac{\tau^2\omega^2}{2}\right)\sin\varphi.
    \end{split}
\end{equation}

\section{CORRELATIONS}
In the absence of amplitude-to-phase coupling mechanisms inside the laser, one would normally expect that timing jitter noise, Schawlow-Townes noise, and shot noise are statistically independent \cite{HausMecozzi}. Nevertheless, many mode-locking mechanisms effectively rely on conversion of self-phase modulation into self-amplitude modulation \cite{APM}. Therefore, such coupling mechanisms may give rise to additional noise \cite{SPIE}. Provided knowledge of the coupling coefficients, the resulting increase in noise can be accounted for. Consequently, the noise estimates in this article and previous publications should be taken as a lower limit. In the absence of such coupling effects, our noise model is expected to yield uncorrelated noise of the three basic observables $\delta\textit{E}_{\text{p}}$, $\delta\textit{t}$, and $\delta\varphi$. Mathematically, one can verify, e.g., the absence of any correlation between timing jitter noise and phase noise by computing the correlation function
\begin{equation}
    \int_{0}^{\infty}\int_{-\pi}^{\pi}\delta\textit{t}(\varphi,\omega)\delta\varphi(\varphi,\omega)\,\text{d}\varphi\,\text{d}\omega. \tag{20a}\label{24}
\end{equation}
Given that $\delta\textit{t}(\omega) = -\delta\textit{t}(-\omega)$ and $\delta\varphi(\omega)$ = $\delta\varphi(-\omega)$, Eq.~(\ref{24}) always vanishes. In other words, $\delta\textit{t}(\omega)$ is an asymmetric function, i.e., $\sum(\delta\textit{t}) = -1$ whereas $\sum(\delta\varphi) = 1$. Along the
same lines, it can be argued that the correlation integral between  $\delta\textit{E}_{\text{p}}$ and $\delta\textit{t}$ as well as that between $\delta\textit{E}_{\text{p}}$ and  $\sum(\delta\varphi)$ vanish as the integrations yield zero, cf. Table $\text{\ref{table1}}$
\begin{equation}
   \int_{0}^{\infty}\int_{-\pi}^{\pi}\delta\textit{E}_{\text{p}}(\varphi,\omega)\delta\varphi(\varphi,\omega)\,\text{d}\varphi\,\text{d}\omega=0 \tag{20b}\label{25}
\end{equation}\\
and
\begin{equation}
   \int_{0}^{\infty}\int_{-\pi}^{\pi}\delta\textit{E}_{\text{p}}(\varphi,\omega)\delta\textit{t}(\varphi,\omega)\,\text{d}\varphi\,\text{d}\omega=0. \tag{20c} \label{26}
\end{equation}
\setcounter{equation}{\theequation+1}
This means that our semi-classical model inherently provides
statistical independence of $\delta\textit{E}_{\text{p}}$, $\delta\textit{t}$, and $\delta\varphi$. This set of variables appears to be complete in describing noise of mode-locked lasers. However, neither chirp variations $\delta\beta$ nor pulse duration variations $\delta\tau$ are anywhere statistically independent from noise in this elementary set. Consequently, chirp variations $\delta\beta$ show the same symmetry as $\delta\varphi$ in Table $\text{\ref{table1}}$, and ASE-induced pulse duration variations $\delta\textit{E}_{\text{p}}$ are strongly correlated with pulse energy fluctuations.
\begin{table}[!hbt]
\centering
    \fontsize{10}{14}\selectfont
\begin{tabular}{c|c|c|c|c}
\toprule
\hline
Obs. & $\propto  \varphi$ & $\propto \omega$ & $\Sigma(\varphi)$ & $\Sigma(\omega)$ \\  \hline \vspace{-1.6cm}
\begin{minipage}{3em}\large{$\delta E_{\rm p}$}\vspace{7em}\end{minipage} &
\centering
\includegraphics[width=0.2\columnwidth]{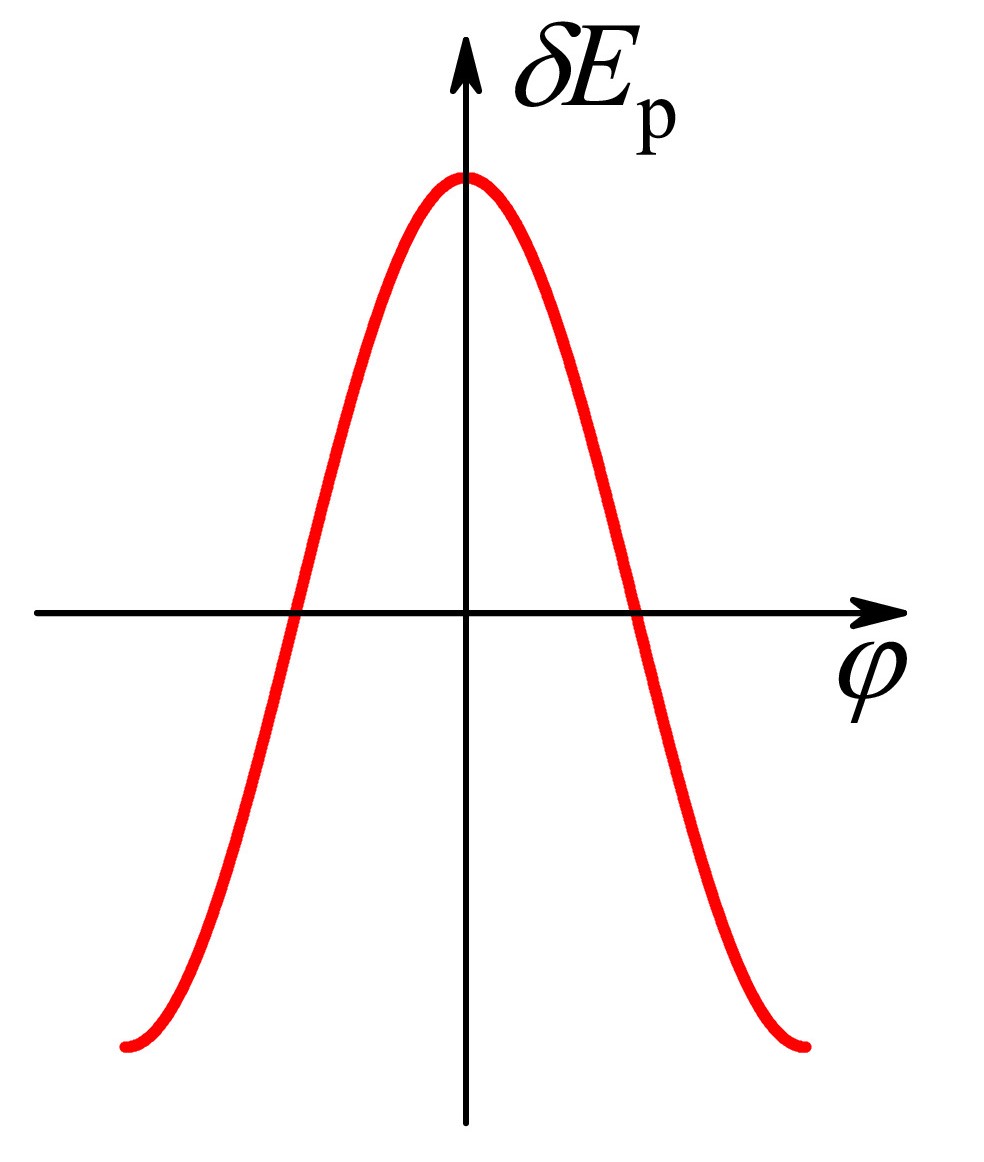} &
\includegraphics[width=0.2\columnwidth]{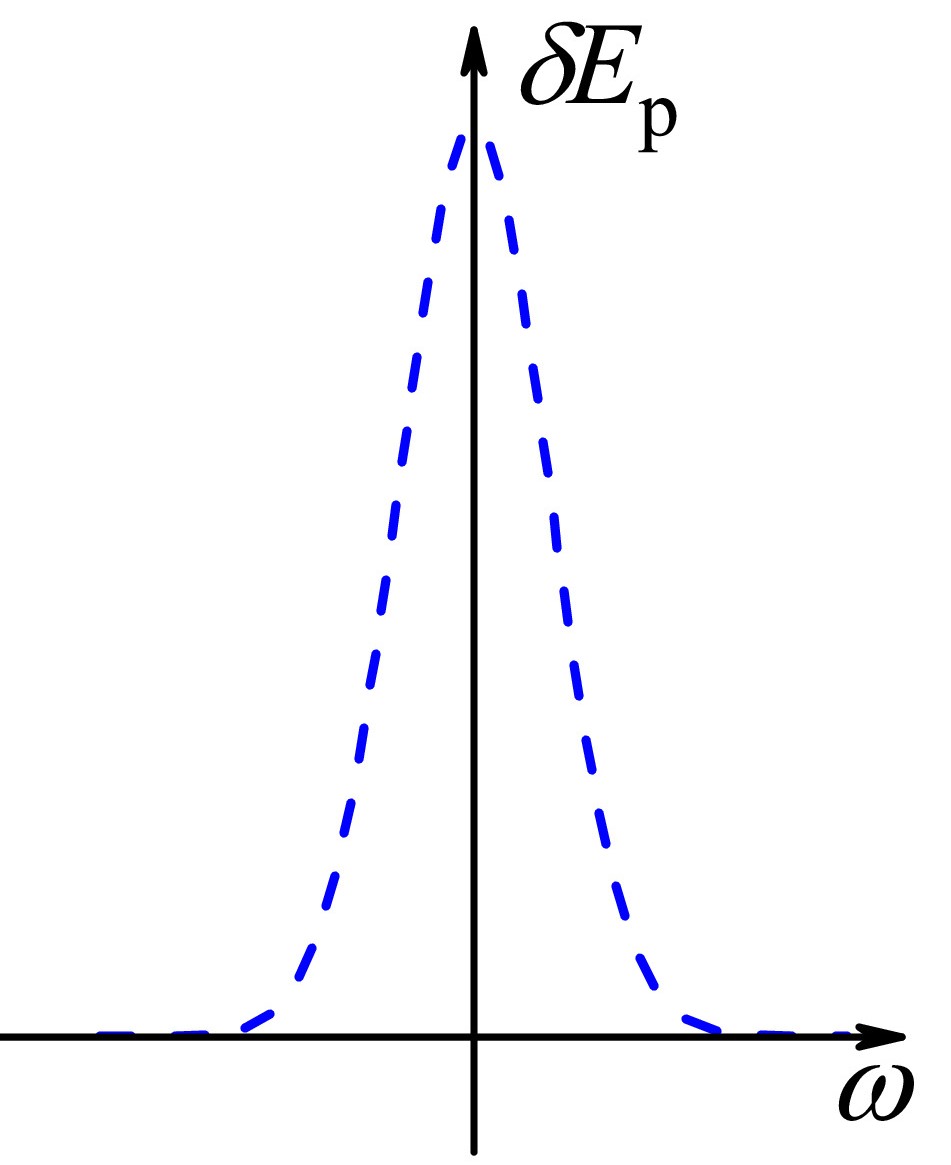} & \begin{minipage}{2em}\huge{$+$}\vspace{4.5em}\end{minipage} & \begin{minipage}{2em}\huge{$+$}\vspace{4.5em}\end{minipage} \\
\hline\vspace{-1.6cm}
\begin{minipage}{3em}\large{$\delta t$} \vspace{7em} \end{minipage}& \includegraphics[width=0.2\columnwidth]{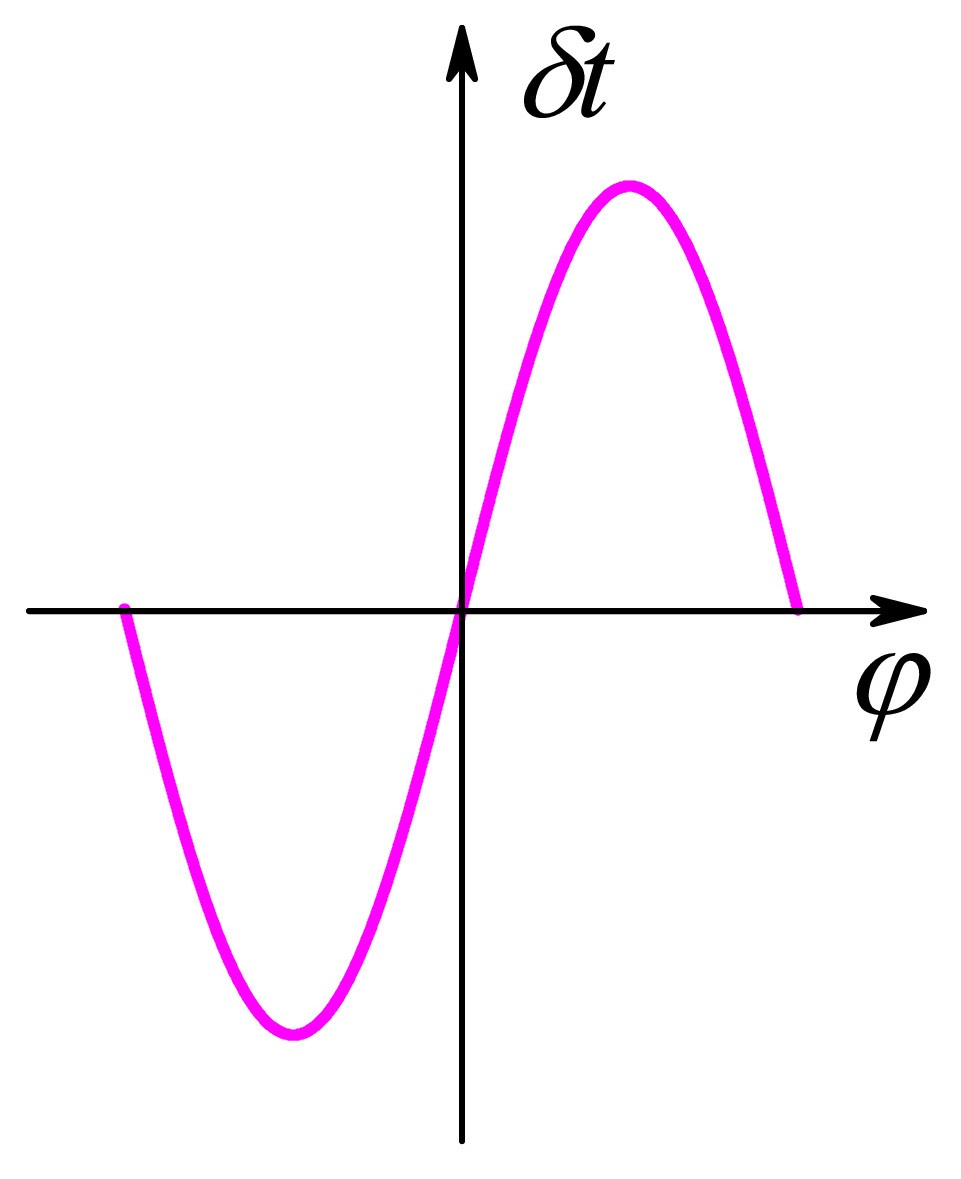} & \includegraphics[width=0.2\columnwidth]{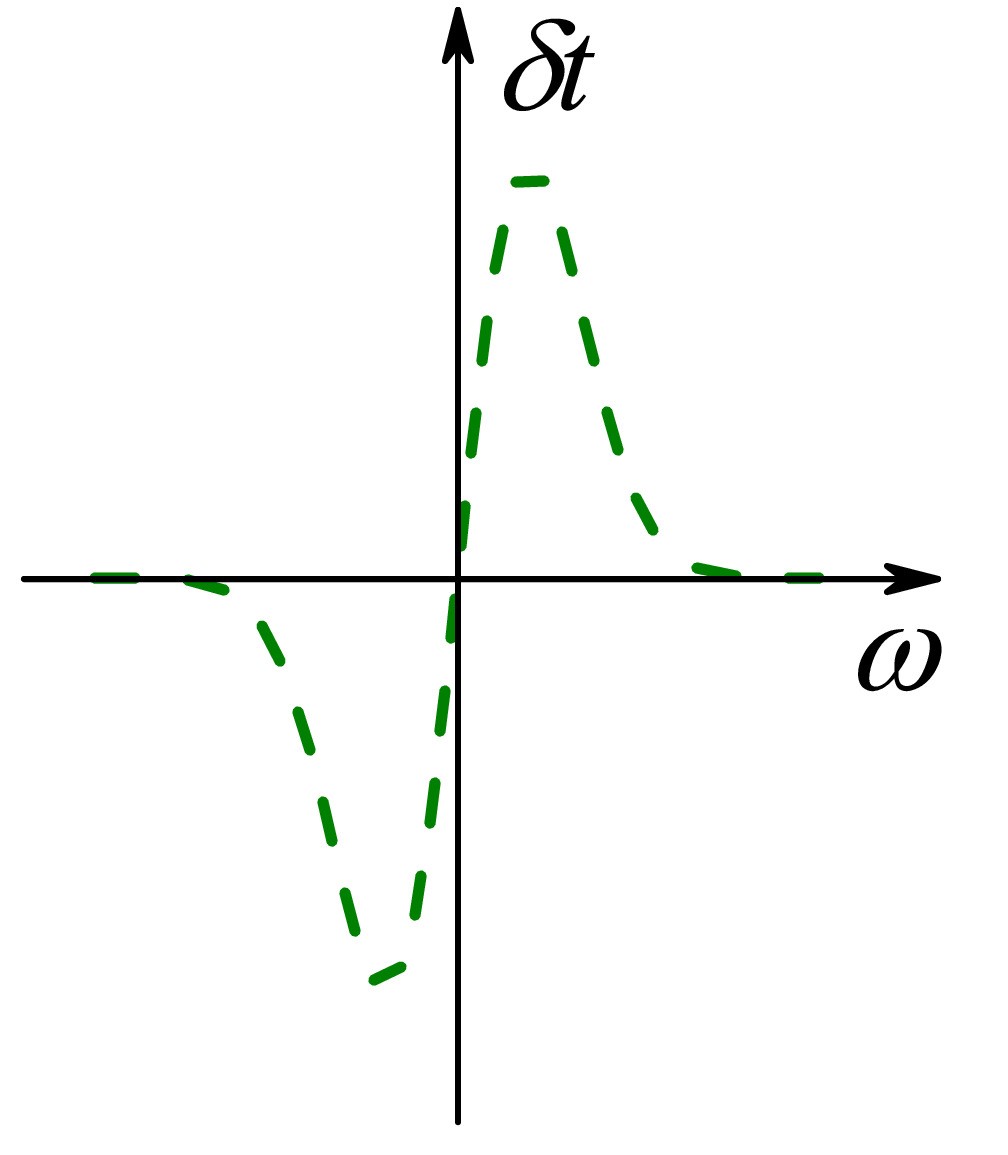} & \begin{minipage}{2em}\huge{$-$}\vspace{4.5em} \end{minipage} & \begin{minipage}{2em}\huge{$-$}\vspace{4.5em}\end{minipage} \\
\hline\vspace{0cm}
\begin{minipage}{3em} \vspace{-3cm}\large{$\delta \varphi$} \vspace{-0.5em}\end{minipage} & \includegraphics[width=0.2\columnwidth]{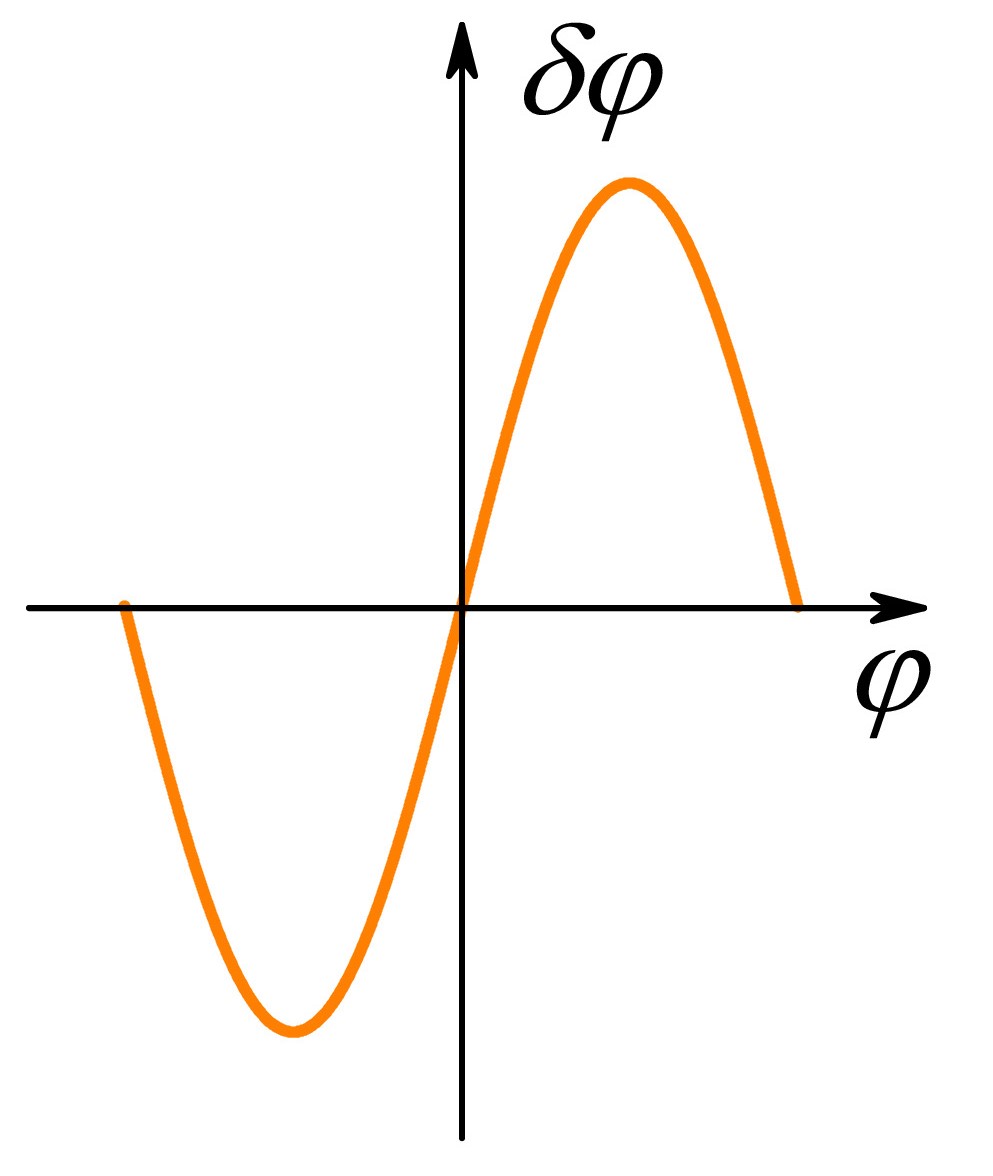} & \includegraphics[width=0.2\columnwidth]{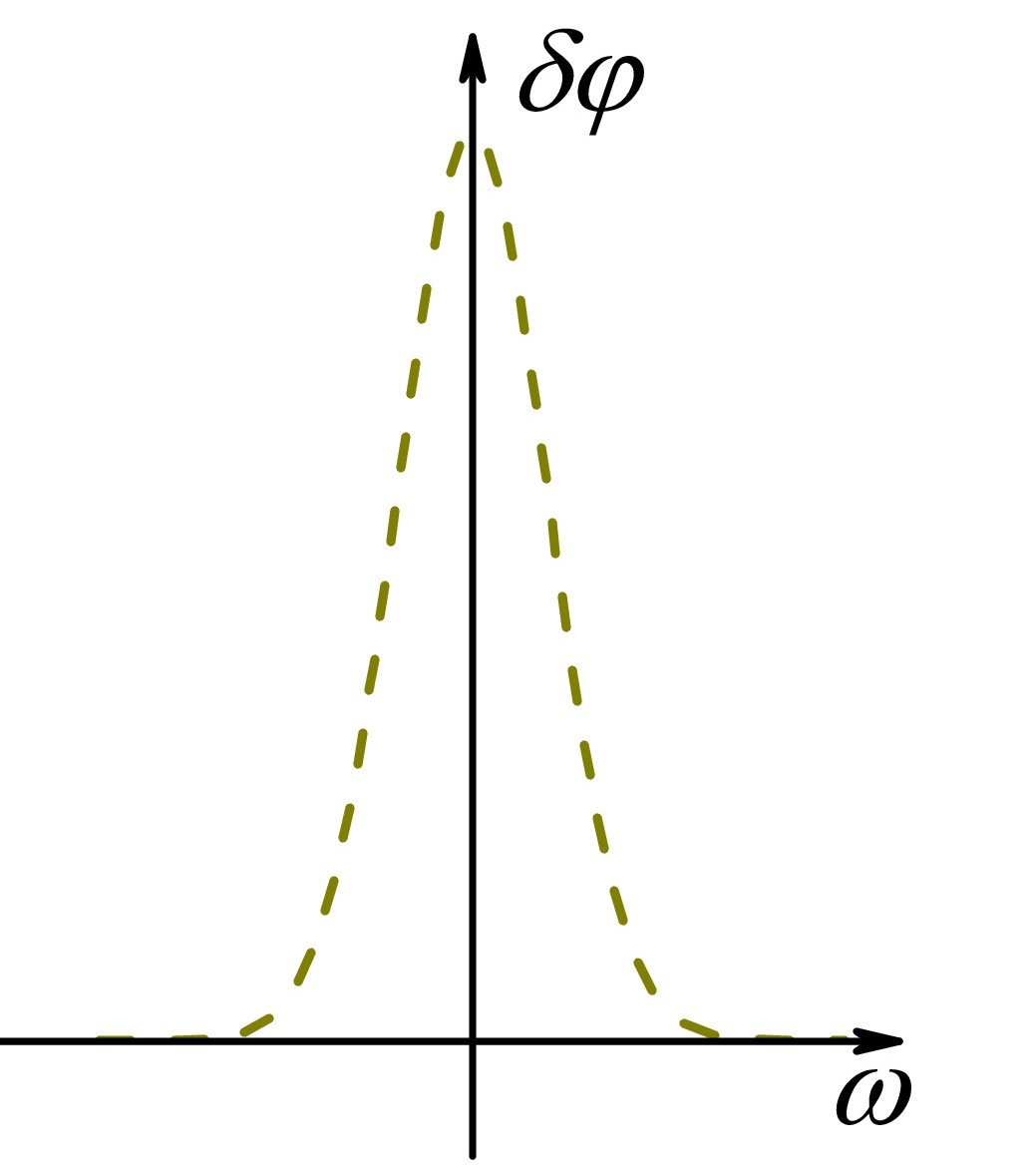} & \begin{minipage}{2em}\vspace{-3cm}\huge{$-$}\vspace{-0.5em}\end{minipage} & \begin{minipage}{2em}\vspace{-3cm}\huge{$+$}\vspace{-0.5em}\end{minipage} \\
\hline
\bottomrule
\end{tabular}
\caption{Symmetries of the underlying stochastic functions in Eqs.~(\ref{11}-\ref{18}) Symmetries are indicated by the $\Sigma$ operator, indicating symmetric behavior with either a $+$ or $-$. The most important conclusion of this diagram is that the mutual product of the symmetries $\Sigma(X_{\omega})\Sigma(Y_{\omega })\Sigma(X_\varphi)\Sigma(Y_\varphi)\equiv -1$, independent of choice of $X$ or $Y$ within $\delta\varphi$, $\delta\tau$, and $\delta E_{\rm p}$ with $X \ne Y$. \label{table1}}
\end{table}

\section{SOLVING THE STOCHASTIC EQUATIONS}
In order to extract shot-to-shot jitters from the above expressions for $\delta\textit{E}_{\text{p}}$, $\delta\textit{t}$ and $\delta\varphi$, it is necessary to compute pertinent expectation values \cite{Drummond}
\begin{equation}\label{27}
    \left<x\right>= \sqrt{\frac{\int_{0}^{\infty}\int_{-\pi}^{\pi}\rho(\varphi,\omega)\left|x(\varphi,\omega)\right|^2
    \,\text{d}\varphi\,\text{d}\omega}{\int_{-\pi}^{\pi}\int_{0}^{\infty}\rho(\varphi,\omega)\,\text{d}\varphi\,\text{d}\omega}}
    \end{equation}
from the above expressions, where $\rho(\varphi)$ and $\rho(\omega)$ describe the statistical distribution of the ASE. For the spectral dependence of $\rho$, we assume a Gaussian distribution. The $\varphi$ dependence is rather trivial as the phase distribution is univariate, i.e., $\rho(\varphi)$ = const. and
\begin{equation}\label{28}
    \sqrt{\frac{\int_{-\pi}^{\pi}\sin^2\varphi\,\text{d}\varphi}{2\pi}}= \sqrt{\frac{\int_{-\pi}^{\pi}\cos^2\varphi\,\text{d}\varphi}{2\pi}}=\frac{1}{\sqrt{2}}.
\end{equation}

For computing the $\omega$ integrals, one has to make explicit assumptions about the fluorescence spectrum and the pulse duration. Assuming a Gaussian spectrum
\begin{equation}\label{29}
    \rho(\omega)=\exp\left[-\left(\frac{\omega}{\Delta\omega}\right)^2\right],
\end{equation}
one receives
\begin{equation}\label{30}
    \frac{\int_{0}^{\infty}\exp\left[-\left(\frac{\omega}{\Delta\omega}\right)^2-\tau^2\omega^2\right]\text{d}\omega}{\int_{0}^{\infty}\exp\left[-\left(\frac{\omega}{\Delta\omega}\right)\right]\text{d}\omega}=\frac{1}{\left(1+\tau^2\Delta\omega^2\right)^{\frac{1}{2}}}
\end{equation}
and
\begin{equation}\label{31}
    \frac{\int_{0}^{\infty}\omega^2\exp\left[-\left(\frac{\omega}{\Delta\omega}\right)^2-\tau^2\omega^2\right]\text{d}\omega}{\int_{0}^{\infty}\exp\left[-\left(\frac{\omega}{\Delta\omega}\right)\right]\text{d}\omega}=\frac{2\Delta\omega^2}{\left(1+\tau^2\Delta\omega^2\right)^{\frac{3}{2}}},
\end{equation}
where $\Delta\omega$ relates to the fluorescence bandwidth of the medium. Completing the stochastic integration, we then yield
\begin{equation}
    \left<E_{\rm p}\right>=2E_{\rm p}\gamma\frac{1}{(1+\tau^2\Delta\omega^2)^{\frac{1}{4}}}, \tag{26a}\label{32}
\end{equation}
\begin{equation}
    \left<\delta\textit{t}\right>=\sqrt{2}\tau^2\gamma\frac{\Delta\omega}{(1+\tau^2\Delta\omega^2)^{\frac{3}{4}}}, \tag{26b}\label{33}
\end{equation}
and
\begin{equation}
    \left<\delta\varphi\right>=\gamma\frac{1}{(1+\tau^2\Delta\omega^2)^{\frac{1}{4}}}. \tag{26c}\label{34}
\end{equation}
Assuming Fourier limited Gaussian pulses, $\tau \Delta\omega = 4 \ln 2 \approx 2.77$ and we can further simplify Eqs.~(\ref{32}--\ref{34}) to
\begin{equation}
    \left<\delta\textit{E}_{\text{p}}\right> = 1.165 \, E_{\text{p}} \gamma, \tag{27a}\label{35}
\end{equation}
\begin{equation}
    \left<\delta\textit{t}\right> = 0.775 \, \tau \gamma \tag{27b}\label{36}
\end{equation}
and
\begin{equation}
    \left<\delta\varphi\right> = 0.582 \, \gamma. \tag{27c}\label{37}
\end{equation}
\setcounter{equation}{\theequation+1}
For the case of nearly transform-limited pulses, these computations relate fluctuations of the major pulse observables directly to the field ratio $\gamma$.
As $\tau^2 \Delta\omega^2$ enters in the denominator of the above expressions, ASE noise will decrease with increasing chirp. Nevertheless, this is only a weak effect, and the exact numerical factors also depend on the exact choice of pulse shape and spectrum. This concludes computation of ASE-induced fluctuations of pulse parameters, which are compared with existing literature in Section \ref{Verification}. Prior to doing so, we continue by computing parameters, for which no analytic ASE noise estimates exist in the current literature. Most importantly, this includes the carrier-envelope phase, which has only been treated numerically so far \cite{Paschotta}. Moreover, we are going to compute chirp and pulse duration fluctuations.

\section{THE CARRIER-ENVELOPE PHASE}
Mode-locked lasers emit a periodic waveform at a repetition rate $\textit{f}_{\text{rep}}$. Mode-locking is enforced by the presence of an active or passive absorber mechanism. In the absence of phase-locking mechanism, intracavity dispersion causes quick dephasing of the oscillating laser modes. A fixed phase relation between the oscillating modes then converts the non-equidistant cold-cavity modes of the laser into an absolutely equidistant comb with spacing $\textit{f}_{\text{rep}}$. The periodic waveform propagates at the group velocity $\textit{v}_{\text{gr}}$ = ($\text{d}\omega$/$\text{d}\textit{k}$). Here $\omega$ is the (angular) frequency and $\textit{k}$ is the wave number. Relating  $\textit{v}_{\text{gr}}$ to the refractive index $\textit{n}$($\omega$) yields \cite{Telle,Helbing}
\begin{equation}
    \frac{1}{\textit{v}_{\text{gr}}} = \frac{1}{\textit{c}}\left(\frac{\text{d}\textit{n}(\omega)}{\text{d}\omega}\omega+\textit{n}(\omega)\right).
     \label{1}
\end{equation}
In contrast, the carrier propagates at the phase velocity $\textit{v}_\varphi$ = $\omega$/\textit{k} = \textit{c}/\textit{n}. The difference between group and
phase differences gives rise to the group-phase offset (GPO)
\begin{equation}
    \Delta\varphi_{\text{GPO}} = \int_{0}^{L}\omega\left(\frac{1}{\textit{v}_\varphi}-\frac{1}{\textit{v}_{\text{gr}}}\right)\text{d}\textit{z}
    \label{2}
\end{equation}
per roundtrip through a cavity with length \textit{L}. For the typical example of a few-cycle Ti:sapphire laser with
a 2.3 mm crystal length and 80 MHz repetition rate \cite{Sutter}, the difference between phase and group velocities amounts to a total of about 120 cycles and 20 cycles in the laser crystal and air path, respectively. Additional
contributions may come from chirped mirrors. It is important to realize here that only fractional number of cycle shifts matter, i.e.,  $\Delta\varphi_{\text{GPO}}$ mod 2$\pi$. From Eq.~$\eqref{2}$, we can now determine the carrier envelope-offset (CEO) frequency
\begin{equation}
    \textit{f}_{\text{CEO}} = \frac{1}{2\pi}\textit{f}_{\text{rep}}(\Delta\varphi_{\text{GPO}}\  \text{mod}\ 2\pi).
    \label{3}
\end{equation}
Fourier transforming the repetitive waveform into the spectral domain yields an equidistant comb with spacing
$\textit{f}_{\text{rep}}$ with offset $\textit{f}_{\text{CEO}}$ at the origin. Individual comb
frequencies $\textit{f}_{m}$ are given by
\begin{equation}
    \textit{f}_{m} = \textit{f}_{\text{CEO}}+\textit{m}\textit{f}_{\text{rep}},\ \textit{m}\in\mathbb{N}_{0}.
    \label{4}
\end{equation}
Furthermore, one can immediately see from Eq.~$\eqref{2}$ that
both Schawlow-Townes noise as well as timing jitter noise
contribute to CEP noise. While Schawlow-Townes noise
causes a phase retardation, timing jitter noise enters via
the group velocity term in Eq.~$\eqref{2}$. In addition, the latter mechanism also affects the comb spacing in Eq.~$\eqref{4}$.
Depending on the relative strength of the two noise contributions, different parts of the comb are more or less susceptible to ASE. If Schawlow-Townes noise is small
compared to timing jitter, modes in the vicinity of \textit{m} = 0
are least susceptible to ASE noise. In the elastic tape
picture of the frequency comb, one therefore seeks the
frequency that experiences the least perturbation due to
ASE \cite{rubberband}. This frequency is usually referred to as the
fixed point of the comb. Previous investigations based
on numerical simulations \cite{Paschotta3} indicated that this fixed
point is near to the carrier frequency $\omega_{0}$.

Exploiting the statistical independence of $\delta\textit{t}$ and $\delta\varphi$, the carrier-envelope phase noise is now simply written as
\begin{equation}\label{38}
\begin{split}
    \left<\delta\varphi_{\text{CE}}\right>&=\sqrt{\left<\delta\varphi\right>^2+\omega_0^2\left<\delta\textit{t}\right>^2}\\&
    \approx 0.775 \, \omega_0 \tau \gamma.
\end{split}
\end{equation}
This shows quite clearly that the dominant source of ASE induced CEP noise is timing jitter and not Schawlow-Townes noise, which agrees with the numerical simulations of Paschotta {\em et al.} \cite{Paschotta}.

\section{CHIRP AND PULSE DURATION}

Let us now conclude by solving the
stochastic function for the pulse duration
\begin{equation}\label{39}
\begin{split}
       \left<\delta\tau\right>&=
       \sqrt{\frac{\int_{0}^{\infty}\int_{-\pi}^{\pi}\rho(\omega)\left|\delta\tau\right(\varphi,\omega)|^2\,\text{d}\varphi\,\text{d}\omega}
       {2\pi\int_{0}^{\infty}\rho(\omega)\,\text{d}\omega}}\\&
       = 8 \sqrt{2} \gamma \tau \left(\frac{1}{\sqrt{8+\tau^2 \Delta\omega^2}} - \frac{2}{(4+\tau^2 \Delta\omega^2)^{3/2}} \right) \\&
       \approx 2.2 \, \tau \gamma.
       %
\end{split}
\end{equation}
Comparing with Eq.~(\ref{36}), it is interesting to note that pulse duration variations are about a factor three stronger than the timing jitter itself. Moreover, these variations of duration are not completely uncorrelated to Eqs.~(\ref{35}-\ref{37}) within the framework of our model. For the chirp jitter, we yield in a similar fashion
\begin{equation}\label{40}
    \begin{split}
    \left<\delta\beta\right>&=
    \sqrt{\frac{\int_{0}^{\infty}\int_{-\pi}^{\pi}\rho(\omega)\left|\delta\beta(\varphi,\omega)\right|^2\,\text{d}\varphi\text{d}\omega}
    {2\pi\int_{0}^{\infty}\rho(\omega)\,\text{d}\omega}}\\&
    =\frac{2\gamma}{\tau^2}\frac{\left(1+2\tau^4\Delta\omega^4\right)^\frac{1}{2}}{\left(1+\tau^2\Delta\omega^2\right)^\frac{5}{4}}\\&
    \approx \frac{1.464}{\tau^2} \gamma.
    \end{split}
\end{equation}

\section{PULSE CONTRAST AND COHERENCE} \label{sec:pedestal}
\begin{figure}[!hbt]
    \centering
    \includegraphics[width=0.8 \linewidth]{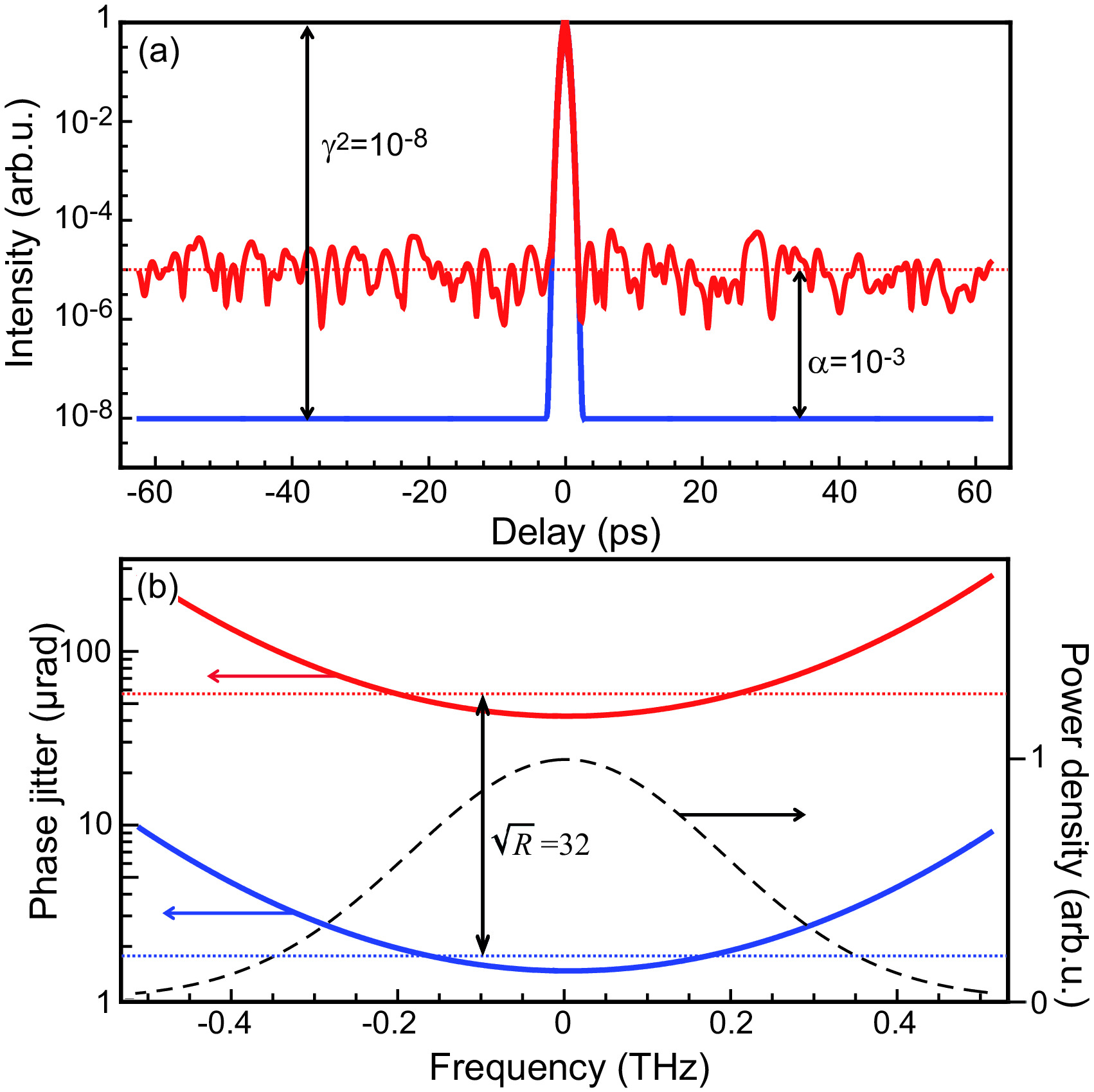}
    \caption{Numerically computed example showing (a) resulting pulse contrast and (b) spectral phase jitter for $\gamma=10^{-4}$, $M=264$, $\alpha=0.1\%$. Blue curves: interference of single random field with amplitude $\sqrt{M} |\mathcal{E}_\phi$ according to Eq.~(\ref{eq:effective}). Red curves: correct modeling of ASE build-up using Eq.~(\ref{eq:fullmodel}). Dotted line in (a): expected ASE background according to Eq.~(\ref{eq:contrast}). Dashed curve in (b): spectral power density. Dotted line: average phase noise. Ensemble size: $R=1000$. Number of iterations to build up the ASE: 10,000. Ensemble averaging yields $\sqrt{R}$ times higher spectral phase noise than single-shot variation.}
 \label{fig:coherence}
\end{figure}

Pulse contrast is usually a topic that is discussed in the context of chirped-pulse amplification (CPA) \cite{Wang,Ricci}. Nevertheless, all CPA laser system start with a mode-locked oscillator, and the oscillator itself already exhibits a limited pulse contrast due to ASE. Given the rather low pulse energies of oscillators, this contrast is certainly hard to measure \cite{Braun}. At the same time, the oscillator pulse contrast imposes a limit for the contrast in subsequent amplification changes. Let us assume that spontaneous emission adds a total of $M$ random photons per roundtrip to the coherent field, where $M$ is again the number of longitudinal modes of the laser. If there is no previous background noise in the cavity the spontaneous emission field strength will grow with $\sqrt{R M}$ for delays that are much larger or shorter than the pulse duration. Here $R$ is the number of roundtrips. Consequently, the background intensity will grow linearly with $R M$. Therefore, in the absence of saturable absorption, one expects the laser to eventually drop out of mode-locking. Nevertheless, in order to stabilize mode-locking, one requires an intracavity saturable absorption mechanism with modulation depth $\alpha$. If sufficiently fast, this absorber imposes additional losses $\alpha$ to the ASE background. As a consequence the field strength of the emitted ASE background decreases by a factor $\sqrt{1-\alpha}$ roundtrip after roundtrip. Consequently, the intensity of an emitted ASE wavepacket drops according to $1-\alpha$ every roundtrip. This allows us to employ the geometric series to compute the resulting intensity in the ASE background
\begin{equation}
|\mathcal{E}_{\rm ASE}|^2 = \sum_{j=0}^\infty (1-\alpha)^{j} M |\mathcal{E}_\phi|^2 = \frac{1}{\alpha} M |\mathcal{E}_\phi|^2
\end{equation}
with resulting intensity contrast
\begin{equation} \label{eq:contrast}
\chi = \frac{M |\mathcal{E}_\phi|^2}{\alpha |\hat{\mathcal{E}}|^2} = \frac{\gamma^2}{\alpha},
\end{equation}
cf.~Fig.~\ref{fig:coherence}. Among the many laser parameters that we discussed so far, the contrast is probably the most accessible quantity. Braun {\em et al.} measured a contrast $\chi \approx 10^{-7}$ for a Kerr-lens mode-locked Ti:sapphire laser \cite{Braun}. This measurement would indicate an effective modulation depth of the Kerr-lensing mechanism well below a percent, see Table \ref{tablex}. Lasers that were mode-locked with a semiconductor saturable absorber, in contrast, showed significantly higher contrast \cite{Braun}, which appears to be compatible with the higher modulation depth of such devices.

One can now additionally derive an estimate for the resulting pulse-to-pulse coherence exploiting Eq.~(\ref{37})
\begin{equation} \label{eq:coh}
\Gamma = \frac{\left| \left<  \mathcal{E}_i(t) \mathcal{E}^*_{i+1}(t) \right>_{i,t} \right|}{
\left< \left| \mathcal{E}_i(t) \mathcal{E}^*_i(t) \right| \right>_{i,t}} \approx \cos\left( \left<\delta \varphi \right>\right) \approx 1 - 0.17 \gamma^2.
\end{equation}
This expression can be inverted to find the number of roundtrips after which the coherence $\Gamma$ has degraded to $1/{\rm e}$. Relating the coherence to Schawlow-Townes noise yields the coherence time
\begin{equation} \label{eq:cohtime}
t_{\rm coh}=\frac{3.73}{\gamma^2 f_{\rm rep}}.
\end{equation}
For solid-state lasers, the Schawlow-Townes limit of the coherence time is usually much larger than a second and is completely overruled by technical noise sources. Even for lasers with rather low values of $\gamma$, Eq.~(\ref{eq:cohtime}) is still expected to indicate coherence times in the millisecond range. Nevertheless, the latter definition can be deceiving for lasers with weakly pronounced saturable absorber mechanism as it ignores the build-up of substantial ASE fields in the individual laser modes.
It strongly depends on the application of the mode-locked laser, e.g., as a frequency comb source or to seed an amplifier whether this buildup of noise causes limitations.

\section{A ONE-PHOTON-PER-MODE NOISE MODEL}

In order to illustrate the connection between ASE noise, pulse contrast, and coherence, we ran a series of numerical simulations, see Fig.~\ref{fig:coherence}. For clarity, we used a value of $\gamma=10^{-4}$ and a relatively poor modulation depth of $\alpha=10^{-3}$. With these parameters, we expect a pulse/ASE intensity contrast of $10^{-5}$ [Eq.~(\ref{eq:contrast})]. In a simple-minded approach to simulating this scenario, one would generate a coherent waveform $\mathcal{E}_{\rm coh}$ and add $M$ incoherent ASE fields with modulus $|\mathcal{E}_\phi|$ each for every roundtrip.
\begin{equation} \label{eq:naive}
\mathcal{E}_k = \mathcal{E}_{\rm coh}(t) + \sum_{j=1}^M \mathcal{E}_\phi \exp ( i \omega_{M k + j} t + i \phi_{M k + j} ).
\end{equation}
Here the random phases $\phi_j$ are univariate, and the random frequencies $\omega_j$ follow a Gaussian distribution with bandwidth $\Delta\omega$. Obviously, this approach is numerically not overly efficient, and it ignores the build-up of ASE photons inside the cavity. One can make this approach more efficient by letting
\begin{equation} \label{eq:effective}
\mathcal{E}_k = \mathcal{E}_{\rm coh}(t) + \sqrt{M} \mathcal{E}_\phi \exp ( i \omega_k t + i \phi_k ),
\end{equation}
using a Gaussian distribution $\Delta\omega / \sqrt{M}$ of the random frequencies $\omega_k$ instead. In order to simulate the ASE buildup inside the cavity, we accumulate cavity roundtrips according to
\begin{eqnarray}
\mathcal{E}_0^{\rm (ASE)} & = & \mathcal{E}_0 \nonumber \\
\mathcal{E}_k^{\rm (ASE)} & = & \sqrt{1-\alpha} \mathcal{E}_{k-1}^{\rm (ASE)} + \mathcal{E}_k. \label{eq:fullmodel}
\end{eqnarray}
Heterodyning these fields with the coherent field $\mathcal{E}_{\rm coh}(t)$, one arrives at the scenario depicted in Fig.~\ref{fig:coherence}(a). As expected from Eq.~(\ref{eq:contrast}), the resulting contrast in our simulation is nearly exactly $10^{-5}$ and exceeds the contrast expected from the simple-minded approach Eq.~(\ref{eq:naive}) by 3 orders of magnitude.

Another interesting aspect of our simulations is revealed in Fig.~\ref{fig:coherence}(b). Regardless of using Eq.~(\ref{eq:effective}) or (\ref{eq:fullmodel}), there is a pronounced spectral dependence of the resulting phase jitter that follows $1/|\tilde{\mathcal{E}}(\omega)|$. While this may not appear overly surprising, it was common practice to assume flat spectral noise for simulating coherence properties in supercontinuum generation \cite{Dudley}. More recently, it has become clear that this assumption is too simplistic as it actually underestimates resulting decoherence effects \cite{Frosz}. It therefore appears important to include an assumption on $|\tilde{\mathcal{E}}(\omega)|$ into the simulation of quantum noise effects in supercontinuum generation to correctly explain the observed decoherence effects.

\section{VERIFICATION} \label{Verification}
\begin{table}
\centering
    \fontsize{10}{12}\selectfont
\begin{tabular}{l|c|c|c}
    \toprule
    \hline
    Parameter & & Eq. & Expression  \\
    \hline
    \multicolumn{4}{l}{Primary parameters} \\
    \hline
    Pulse energy & $\left<\delta E_{\rm p}\right>$ & \eqref{35} & $1.165 \gamma E_{\rm p}$ \\
    Delay & $\left<\delta t\right>$ & \eqref{36} & $0.775 \gamma \tau$ \\
    Phase & $\left<\delta\varphi\right>$ & \eqref{37} & $0.582 \gamma$ \\
    \hline
    \multicolumn{4}{l}{Secondary parameters} \\
    \hline
    Carrier-envelope phase & $\left<\delta \varphi_{\rm CE}\right>$ &  \eqref{38} &  $0.775 \gamma \tau \omega_0 $ \\
    Duration & $\left<\delta \tau\right>$ & \eqref{39} &  $2.2 \gamma \tau  $ \\
    Chirp & $\left<\delta \beta\right>$ & \eqref{40} & $1.464 \gamma / \tau^{2}$ \\
    \hline
    \multicolumn{4}{l}{Coherence parameters} \\
    \hline
    Contrast & $\chi$ &  \eqref{eq:contrast} &  $\gamma^2/\alpha$ \\
    Decoherence & $1-\Gamma$ & \eqref{eq:coh} &  $0.17 \gamma^2 $ \\
    Coherence time & $t_{\rm coh}$ & \eqref{eq:cohtime} & $3.73 / \gamma^2 f_{\rm rep}$ \\
    \bottomrule

\end{tabular}
\vspace{0cm}
\caption{ASE-induced single-shot fluctuations of mode-locked laser variables and resulting coherence properties.}\label{tablex}
\end{table}

Let us now verify our formalism and plug in typical numbers from a few-cycle Ti:sapphire oscillator \cite{Sutter} with 80 MHz repetition rate, 6\,fs pulse duration and intracavity pulse energy of 50 nJ. For the Ti:sapphire gain medium, we assume $\Delta\omega$ = 4.3 $\times$ 10$^{14}$ rad/s and $\omega_0$ = 2.4 $\times$ 10$^{15}$ rad/s. With these numbers, we compute a number of coherent intracavity photons $\textit{N} = 2 \times 10^{11}$. Assuming one added ASE photon per mode \cite{Dudley,Frosz}, one expects an accumulated number $\textit{M} = 0.85 \times 10^{6}$ of incoherent photons per cavity roundtrip of the coherent field.
Using Eqs.~(\ref{6a} -- \ref{11a}), we then compute a resulting field ratio
\begin{equation}\label{46}
 \gamma = \sqrt{\frac{\sqrt{\pi}\tau\textit{M}f_{\text{rep}}}{N}} = \sqrt{\frac{\tau \Delta\omega}{2 \sqrt{\pi} N}}.
 \end{equation}
With the above parameters, we estimate $\gamma = 1.9 \times 10^{-6}$ for a few-cycle Ti:sapphire laser. For all of the following verifications with existing literature, this estimate is the key parameter.

Among the discussed noise mechanisms, shot noise probably provides the most direct verification of our considerations. Using Eq.~(\ref{35}), we compute a resulting pulse energy jitter $\left< E_{\rm p} \right> = 11\,$pJ. This has to be compared with standard shot noise for non-squeezed light. In this situation one expects that the variance of the pulse energy equals the average energy, which leads to the following equation for the photon number $\textit{N}$ \cite{Paschotta}
\begin{equation} \label{41}
    \left(\frac{\delta\textit{E}_{\text{p}}}{\textit{E}_\phi}\right)^2 = N.
\end{equation}
This relation can be rewritten to yield
\begin{equation}\label{42}
  \delta\textit{E}_{\text{p}}=\sqrt{N}\textit{E}_\phi,
\end{equation}
which provides the identical value as our Eq.~(\ref{35}) when we plug in $N=2 \times 10^{11}$. It should be emphasized that the effective number of modes $M$ (or the bandwidth $\Delta\omega$) has to be carefully chosen to avoid any discrepancy here. Standard shot noise therefore appears the best way to obtain a reliable estimate for the effective number of modes $M$.

The verified value of $\gamma = 1.9 \times 10^{-6}$ can now be plugged into Eq.~(\ref{37}) to yield an estimate of the pulse-to-pulse phase jitter, i.e., $\left<\delta\varphi\right>=1.1\,\mu$rad. This is to be compared with established equations in the literature, namely Eq.~$\eqref{6}$ from Ref. \cite{Paschotta2}
\begin{equation}\label{47}
    \delta\varphi=\sqrt{\frac{E_\phi}{2E_{\text{p}}}},
\end{equation}
which yields $\delta\varphi = 1.6\,\mu$rad. The slight discrepancy may be explained by the fact that one usually assumes a continuous-wave laser in the derivation of the Schawlow-Townes linewidth.

Moreover, we estimate a pulse-to-pulse timing jitter $\left<\delta t\right> =  8.8 \times 10^{-21}$\,s. For comparison, we use Eq.~$\eqref{8}$ from Ref. \cite{Paschotta}
\begin{equation}\label{48}
    \delta\textit{t}\approx\sqrt{\frac{0.2647}{N}}\tau,
\end{equation}
and end up with 6.9 $\times$ $10^{-21}$ s, i.e., again a reasonable agreement. We can now proceed to compute an estimate
for the single-shot CEP noise $\left<\delta\varphi_\text{CE}\right>$ = 21 $\mu$rad. This number is obviously orders of magnitude below the best reported time-averaged CEP measurements, which lie in the single milliradian range \cite{Liao,Lemons}. For estimation of the resulting effect on larger time scales, we define the phase noise density
\begin{equation}\label{49}
    S_{\varphi, \text{CE}}=f_{\text{rep}}\left<\delta\varphi_{\text{CE}}\right>^2,
\end{equation}
which equals to 0.035 rad$^2\cdot$Hz. Using this value, one can, e.g., estimate the ASE-limited CEP jitter on typical stabilization servo bandwidths of 10 kHz as 3\,mrad, which is reasonably close to the best demonstrated CEP stabilization performance \cite{Liao,Lemons}.

Let us now conclude this section by computing the ASE-limited single-shot chirp and pulse duration variation. From Eq.~$\eqref{40}$ we compute  $\left<\delta\beta\right> = 7.7 \times 10^{22} \ \text{rad/s}^2$, which is more accessibly expressed in units of carrier frequency per pulse duration, i.e., $\left<\delta\beta\right> = 2 \times 10^{-7} \omega_0/\tau$. This leads to stochastic carrier frequency variations in the MHz range within the pulse duration. Using Eq.~$\eqref{39}$, we
finally estimate $\left<\delta\tau\right>= 2.5 \times 10^{-20} $s which is more than 5 orders of magnitude below the assumed pulse duration.

A problematic issue is the determination of the modulation depth $\alpha$ of the Kerr lens inside a Ti:sapphire laser. Theoretical calculations \cite{Lin} predict values $\alpha>0.1$, which are expected to lead to contrast values $\chi \ll 10^{-10}$ and are beyond current measurement capabilities. A more realistic estimate may therefore be $\alpha \approx 10^{-3}$, which still surpasses contrast measurements \cite{Braun} by an order of magnitude.

While the resulting noise mechanisms do not impose any apparent limitation on Ti:sapphire lasers or other lasers with rather high intracavity power, the factor $\gamma$ immediately enters into all jitters in Eqs.~(\ref{35}--\ref{40}). It is therefore clear that ASE becomes a much more significant perturbation for lasers with low intracavity power, long pulses, and short cavities We therefore analyze two examples \cite{Faist,SML} of reported mode-locking of semiconductor lasers and compare them with the Ti:sapphire laser in Table $\text{\ref{table2}}$. Comparing ASE-induced shot-to-shot variations of the pulse
duration $\left<\delta\tau\right>$ make it rather clear that in particular for
the case of a mode-locked vertical external-cavity surface emitting laser (VECSEL), there is an obvious need for a strong mode-locking mechanism to counteract quantum noise effects. Moreover, temporal characterization of these lasers was indicative of a rather strong background \cite{Faist,SML} of $3$ -- $30\%$, and coherence characterization showed a possible coherence degradation \cite{Burghoff}. If we take the reported numbers seriously then they can only be explained by a rather weak mode-locking mechanism with an effective modulation depth $\alpha ~ 10^{-6}$ or lower. Mode-locked semiconductor lasers are therefore expected to show measurable quantum noise effects that appear much more accessible than the feeble effects in all-solid-state lasers.
\begin{table}
\centering
    \fontsize{10}{12}\selectfont
    \label{tab:verify}
\begin{tabular}{c|c|c|c|cc}
    \toprule
    \hline
    Parameter & Eq. & Ti:Sa & QCL & VECSEL \\
     &  & \cite{Sutter} & \cite{Faist,Burghoff} & \cite{SML} \\
    \hline
    $\omega_0$\ ($10^{15}$\,\text{rad/s})& &2.4&0.23&1.91\\
    \hline
    $\Delta\omega$\ ($10^{12}$\,\text{rad/s})& &350&19&2.7\\
    \hline
    $\tau$\ (\text{fs})& &6&13400&930\\
    \hline
    $N$ & &2\ $\times$\  $10^{11}$&$10^{9}$&$10^{7}$\\
    \hline
    $M$ & &0.85\ $\times$\ $10^{6}$&400&2000\\
    \hline
    $\alpha $ & & 0.001 &$2 \times 10^{-7} {}^\dagger$ & $2 \times 10^{-6} {}^\dagger $ \\
    \hline
    $\gamma$\ ($10^{-6}$)&\eqref{46} &{\bf 1.9}&{\bf 260}&{\bf 260}\\
    \hline
    $\left<\delta\varphi\right>$\ ($\mu$\text{rad})&\eqref{37}&1.1&150&150\\
    \hline
    $\left<\delta\textit{t}\right>$\ (\text{as})&\eqref{36}&0.009&2700&185\\
    \hline
    $\left<\delta\tau\right>$\ (\text{as})&\eqref{39}&0.025&7500&550\\
    \hline
    $\left<\delta\beta\right>(\tau/\omega_0)$&\eqref{40}&$2 \times 10^{-7}$&$10^{-7}$&$2\times 10^{-7}$\\
    \hline
     $\chi$ & \eqref{eq:contrast}& $4 \times 10^{-9}$ &$0.3 {}^\dagger$ & $0.03 {}^\dagger$ \\
    \hline
    $\delta\varphi_{\text{CE}}$\ (\text{mrad})&\eqref{38}&0.02&600&350\\
    \hline
     $1-\Gamma$\ & \eqref{eq:coh}&$10^{-12}$&$10^{-8}$&$10^{-8}$\\
    \hline
     $t_{\rm coh}$ & \eqref{eq:cohtime}& $\gg 1$\,s & 8\,ms & 100\,ms \\
     \hline
    \bottomrule
\end{tabular}\vspace{0cm}
\caption{Comparison of the estimated ASE effects and the potential as comb sources for three different laser technologies. The well-established Ti:sapphire laser was used as a reference \cite{Sutter}. Ref. \cite{Faist} discusses a frequency comb based on quantum cascade lasers, and Ref. \cite{SML} claimed self-modelocking of a semiconductor laser. $\dagger$: Values for $\alpha$ for the latter two have been chosen to best explain observed backgrounds in temporal characterization measurements \cite{Braun,Faist,SML} and $\ddagger$ spectral coherence measurements \cite{Burghoff}.} \label{table2}
\end{table}

\section{CONCLUSION}
In summary, we have presented a completely analytic
approach to describe all implications of ASE noise in a mode-locked laser. In our semi-classical model, the coherent repetitive waveform interferes with temporally unlocalized fields at random phase and frequency. The key parameter for the strength of the individual jitters is the field ratio $\gamma$ between
the peak electric field of the coherent waveform and the average field strength of the continuous ASE field. Using the assumption of one emitted ASE photon per roundtrip and laser mode, we calibrate our model with standard shot noise and yield excellent agreement with published equations for timing noise jitter and Schawlow-Townes noise. In the framework of our model, the three elementary jitter mechanisms have been shown to be uncorrelated with each other. Under the assumption of transform-limited Gaussian pulses, we derive simple expressions for all relevant laser parameters, including ASE-induced pulse duration and chirp variations, for which no estimates have been published so far. We further expand our analysis to provide estimates for the spectral and temporal coherence properties of a mode-locked laser. For typical mode-locked solid-state lasers, coherence properties appear to be out of reach for any measurement. Yet, ASE-induced coherence degradation mechanism may be readily observable in mode-locked semiconductor lasers with weak self-amplitude modulation. We further use our formalism to derive an analytical expression for the quantum-limited CEP jitter, which is nevertheless shown as too weak to play any limiting role in currently established fiber-laser and solid-state-laser based comb technology. Again, for lasers with short cavities, low pulse energies, and long pulse duration, ASE-induced CEP jitters may be sizeable and impose severe limitations for their use as frequency comb sources.

Based on our analytic equations, we developed a concise numerical model that enables the investigation of ASE effects in supercontinuum generation or other nonlinear propagation scenarios. Other than previous approaches, this model includes the build-up of a contrast-limiting ASE background in the cavity and correctly considers the spectral dependence of resulting phase jitters. We believe that the materials in this article will prove useful to estimate the severity of ASE effects in a wide range of mode-locked lasers. While we do not see immediate consequences for a lasers that rely on well-established mode-locking mechanisms there is a number of reports on mode-locking or comb formation that still require theoretical explanations for the observed mode-locking effect. Our formalism may also prove useful to judge the suitability of these novel sources for particular applications. Another intriguing aspect is the role of ASE-induced CEP fluctuations for precision frequency metrology. Even though pulse-to-pulse fluctuations appear to be tiny, they may accumulate to a sizeable effect on time scales of minutes or hours. Based on the formalism developed here, the hitherto unexplored influence on ASE on frequency comb measurements will be addressed in future publications.

\section*{ACKNOWLEDGMENT}

We gratefully acknowledge fruitful discussions with John Dudley (Universit\'e Franche-Comt\'e),  J\'er\^ome Faist (ETH Z\"urich), Go\"ery Genty (Tampere University), Sebastian Koke (PTB Braunschweig), R\"udiger Paschotta (RP Photonics), and Markus Pollnau (University of Surrey).
We further acknowledge financial support by Germany's Excellence Strategy within the Cluster Excellence PhoenixD (EXC 2122, Project ID 390833453).

R.L. and C.M. contributed equally to this work.

\end{document}